\documentclass[a4paper]{toledoStyle}
\usepackage{epsfig}

\begin{document}

\title{Radiative Transfer for the $\bf{\cal {FIRST}}$ ERA}

\author{J.\,Trujillo Bueno\inst{1,2}} 

\institute{
  Instituto de Astrof\'\i sica de Canarias, E-38200, La Laguna, Tenerife, Spain
\and
  Consejo Superior de Investigaciones Cient\'\i ficas, Spain.}

\maketitle 


\begin{abstract}

This paper presents a brief overview of some recent advances
in numerical radiative transfer, which may help the molecular astrophysics 
community to achieve new breakthroughs in the
interpretation of spectro-(polarimetric) observations.

\end{abstract}


\section{Introduction}
\label{introd}

The development of novel Radiative Transfer (RT) methods often leads
to important breakthroughs in astrophysical plasma spectroscopy because
they allow the investigation of problems that could not be properly tackled
using the methods previously available. This RT topic is also of great
interest for the ``Promise of FIRST'', since a rigorous interpretation
of the observations will require to carry out detailed confrontations
with the results from NLTE RT simulations in one-, two-, and three-dimensional
geometries. 

The efficient solution of NLTE multilevel RT problems requires
the combination of a highly convergent iterative scheme with a very fast
formal solver of the RT equation. This applies to the case
of unpolarized radiation in atomic lines, to the promising topic of
the generation and transfer of polarized radiation in magnetized plasmas
and to RT in molecular lines.

The ``dream'' of numerical RT is to develop iterative methods
where everything goes as simply as with the well-known $\Lambda$-iteration
scheme, but for which the convergence rate is extremely high. In this  
contribution we present an overview of some iterative methods 
and formal solvers we have developed for RT applications.
Our RT methods are based on Gauss-Seidel
iteration and on the {\it non-linear} multigrid method. 
These new RT developments are of interest because they allow the solution of
a given RT problem with an order-of-magnitude of improvement
in the total computational work with respect to the popular ALI method
(on which most present NLTE codes are based on). Our RT methods
have been succesfully applied to astrophysical problems of
unpolarized radiation in atomic lines (in 1D, 2D and 3D with
multilevel atoms) and also to the transfer of polarized radiation
in magnetized plasmas including anisotropic pumping (Trujillo Bueno, 1999; 2001),
which may be of interest for modelling polarization phenomena in MASERS. 
The case of RT in molecular lines is presented
in an extra contribution at this conference by 
Asensio Ramos, Trujillo Bueno and Cernicharo (2001).
  

\section{RT methods based on Gauss-Seidel iteration}
\label{GS}

The essential ideas behind the iterative schemes
on which our NLTE multilevel transfer codes are based on
can be easily understood by
considering the ``simplest'' NLTE problem: the 
coherent scattering case with a source function given by

\begin{equation}
{\rm S=(1-\epsilon)J\,+\,{\epsilon}\,B},
\end{equation}
with $\epsilon$ the NLTE parameter, J the mean intensity
and B the Planck function. 
The mean intensity at point ``$i$" is the angular average of
incoming (``$in$") and outgoing (``$out$")
contributions. For instance, for a one-point angular quadrature

\begin{equation}
{\rm J_{i}}={\rm J_{i}}^{in}+{\rm J_{i}}^{out}{\approx}{1\over{2}}
({\rm I}_{i}^{in}+{\rm I}_{i}^{out}).
\end{equation}

The well-known $\Lambda-$iteration scheme is to do the following 
in order to obtain the ``${\rm new}$" estimate of the source function
at each spatial grid-point ``$i$":

\begin{equation}
{\rm S_{i}^{{\rm new}}=(1-\epsilon)J_{i}^{{\rm old}}\,+\,\epsilon\,B_{i}},
\end{equation}
where $\rm J_{i}^{{\rm old}}$ is the mean intensity at the grid-point ``$i$"
calculated using the previous known values of the source function (i.e.
using $\rm S_{i}^{{\rm old}}$). 

For a given spatial grid of NP points the formal solution of the transfer
equation can be symbolically represented as

\begin{equation}
{\bf I}_{\Omega}={\vec \Lambda}_{\Omega}\,[{\bf S}]\,+\,{\bf T}_{\Omega},
\end{equation}
where ${\bf T}_{\bf \Omega}$ gives the transmitted specific intensity
due to the incident radiation at the boundary and 
${\vec {\Lambda}}_{\bf \Omega}$
is a NP$\times$NP operator whose elements depend on the optical
distances between the grid-points. Thus, the mean intensity at the grid-point "$i$" would be:

\begin{equation}
{\rm J_{i}=\Lambda_{i,1}S_{1}^{a}+...+\Lambda_{i,i-1}S_{i-1}^{a}+
\Lambda_{i,i}S_{i}^{b}+}
\end{equation}

${\rm +\Lambda_{i,i+1}S_{i+1}^{c}+...+\Lambda_{i,NP}S_{NP}^{c}+T_{i}}$.

$  $

In this last expression ${\Lambda}_{ij}=\sum({\Lambda}_{ij}^{in}+{\Lambda}_{ij}^{out})/2$
(with the sum applied to all the directions of the numerical angular quadrature)
and $a$, $b$ and $c$ are simply symbols that we use as a 
notational trick to indicate
below whether we choose the ``${\rm old}$" or the ``${\rm new}$" values of the source
function. Thus, for instance, the $\Lambda$-iteration method consists
in calculating $\rm J_{i}$ choosing $a=b=c={\rm old}$, 
which gives $\rm J_{i}=J_{i}^{{\rm old}}$ as indicated in Eq. (3).

The Jacobi method, known in the RT literature as the OAB method
(from Olson, Auer and Buchler, 1986), and on which most NLTE codes
are based on, is found by choosing $a=c={\rm old}$, but $b={\rm new}$, which yields

\begin{equation}
{\rm J_{i}=J_{i}^{{\rm old}}\,+\,{\Lambda}_{ii}\,(S_{i}^{{\rm new}}\,-\,S_{i}^{{\rm old}})=
J_{i}^{{\rm old}}\,+\,{\Lambda}_{ii}\,{\delta}S_i}
\end{equation}

In fact, using this expression instead of $\rm J_i^{{\rm old}}$ in Eq. (3)
one finds that the resulting Jacobi iterative scheme is

\begin{equation}
{\rm S_{i}^{{\rm new}}=S_{i}^{{\rm old}}+\delta{S}_{i}},
\end{equation}
with the correction

\begin{equation}
{\rm \delta{S}_{i}={{[(1-\epsilon)J_{i}^{{\rm old}}+{\epsilon}B_{i}-S_{i}^{{\rm old}}]}\over
{[1-(1-\epsilon)\Lambda_{ii}]}}},
\end{equation}
where $\Lambda_{ii}$ is the diagonal element of the $\Lambda$-operator
associated to the spatial grid-point ``$i$". Note that the correction corresponding to the 
slowly convergent $\Lambda$-iteration method is given by Eq. (8), but
with ${\Lambda}_{ii}=0$.

A superior type of iterative schemes are
the Gauss-Seidel (GS) 
based methods of Trujillo Bueno and Fabiani Bendicho (1995), which
can also be suitably generalized to the polarization transfer case
(cf. Trujillo Bueno \& Landi Degl'Innocenti, 1996; Trujillo Bueno and Manso Sainz, 1999).
This type of iterative schemes are obtained by choosing $c={\rm old}$ and $a=b={\rm new}$. This yields

\begin{equation}
{\rm J}_{i}={\rm J}_{i}^{{\rm old}\,{\rm and}\,{\rm new}}\,+\,{\Lambda}_{ii}\,\delta{\rm S}_{i},
\end{equation}
where ${\rm J}_{i}^{{\rm old}\,{\rm and}\,{\rm new}}$ is the mean intensity
calculated using the ``${\rm new}$" values of the source function at
grid-points 1,2,...,$i-1$ and the ``${\rm old}$" values 
at points $i, i+1, i+2$, ...., NP. The iterative correction is given by

\begin{equation}
{\rm \delta{S}_{i}^{GS}={{[(1-\epsilon)J_{i}^{{\rm old}\,{\rm and}\,{\rm new}}+
{\epsilon}B_{i}-S_{i}^{{\rm old}}]}\over
{[1-(1-\epsilon)\Lambda_{ii}]}}}
\end{equation}

It is important to clarify the meaning of this last equation: 

1) First, at point $i=1$
(which we can freely be assigned to any of the two boundaries of the medium
under consideration) use ``old" source function values
to calculate ${\rm J}_{1}$ via a formal solution. Apply Eqs. (10) 
and (7) to calculate ${\rm S}_{1}^{{\rm new}}$.

2) Go to the next point $i=2$ and use ${\rm S}_{1}^{{\rm new}}$ and the 
``old" source-function values ${\rm S}_{j}^{{\rm old}}$
at points $j=2,3,...,$NP to get ${\rm J}_{2}$ via a formal solution.
Apply Eqs. (10) and (7) to calculate ${\rm S}_{2}^{{\rm new}}$.

3) Go to the next spatial point $k$ and use 
the previously obtained ``new" source function
values at $j=1,2,...,k-1$, but the still ``old"
ones at $j=k,k+1,...,$NP to get ${\rm J}_{k}$ via a formal solution
and ${\rm S}_{k}^{{\rm new}}$ as dictated by Eqs. (10) and (7).

4) Go to the next point $k+1$ and 
repeat what has just been indicated in the previous point
until arriving to the other ``boundary point''. Having reached
this stage iniciate again the same process, but choosing
now as ``first point $i=1$'' the above-mentioned ``boundary point''.

The result of what we have just indicated is a pure GS iteration.
Actually, after an {\it incoming} and {\it outgoing} pass we get
two GS iterations! A SOR iteration is obtained by doing the corrections as follows:

\begin{equation}
{\rm \delta{S}_{i}^{SOR}}={\omega}\,{\rm \delta{S}_{i}^{GS}},
\end{equation}
where $\omega$ is a parameter with an optimal value between 1 and 2 
that can be found easily (see Trujillo Bueno and
Fabiani Bendicho, 1995).

Figure 1 shows an example of
the convergence rate of all these iterative methods
applied to a NLTE polarization transfer problem in a stellar model atmosphere
permeated by a constant magnetic field that produces a
Zeeman splitting of 3 Doppler widths.
We point out that the computing
time {\it per iteration} is similar for all these methods and that
matrix inversions are not performed. Thus, the important
point to keep in mind is that our implementation
of the GS method is 4 times faster
than Jacobi (i.e. than the ALI method on which most NLTE codes are based on), 
while our SOR method for radiative transfer applications is 10 times faster.

\begin{figure}[ht]
  \begin{center}
    \epsfig{file=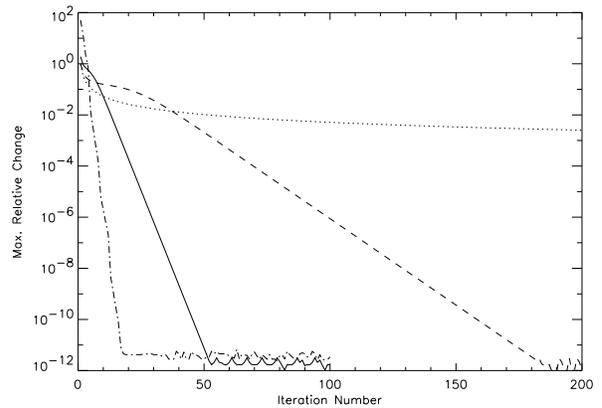, width=8cm}
  \end{center}
\caption{The variation of the maximum relative change
versus the iteration number for several types of iterative methods applied
to the NLTE Zeeman line transfer problem 
discussed by Trujillo Bueno and Landi Degl'Innocenti (1996).
The NLTE parameter $\epsilon=10^{-4}$. Dotted line:
the $\Lambda$-iteration method. Dashed line:
the Jacobi-based ALI method. Solid line: the GS-based method.
Dashed-dotted line: the SOR method.}  
\label{fig5}
\end{figure}

For pedagogical reasons we have chosen here a
NLTE {\it linear} problem in order to explain in simple terms our GS-based
iterative schemes. The generalization to the full {\it non-linear}
multilevel problem can be carried out as indicated
in the Appendix of the paper by Trujillo Bueno (1999). The 
critical point is always to remember that the approximations
one introduces for achieving the required linearity of the statistical equilibrium equations
at each iterative step should treat adequately the {\it coupling} between transitions
and the {\it non-locality} of the problem (see Socas-Navarro \& Trujillo Bueno, 1997).

\section{The non-linear multigrid RT method}
\label{GS}

Our GS and SOR radiative transfer methods
are based, like the Jacobi-like ALI method, on the idea of operator splitting.
Therefore, all of them are characterized by a convergent rate which
{\it decreases} as the resolution of the spatial grid is increased.
As a result, if NP is the number of grid-points in a computational box
of {\it fixed} dimensions, the computing time or
computational work (${\cal W}$) required by the three
previous iterative methods to yield the self-consistent atomic (or molecular)
level populations scales with NP as follows (cf. Trujillo Bueno \& Fabiani Bendicho, 1995):

\begin{itemize}

\item Jacobi-based ALI method $\rightarrow\,$ ${\cal W}{\sim}{\rm NP}^2$

\item Our GS-based RT method $\rightarrow\,$ ${\cal W}{\sim}{\rm NP}^2/4$

\item Our SOR RT method $\rightarrow\,$ ${\cal W}{\sim}{\rm NP}\sqrt{{\rm NP}}$

\end{itemize}

Is there any suitable RT multilevel method characterized by ${\cal W}{\sim}{\rm NP}$?
This would be of great interest for 
3D RT with multilevel atoms where NP$\sim10^6$.
The answer is affirmative. This has been worked out by
Fabiani Bendicho, Trujillo Bueno and Auer (1997) who considered
the application of the {\it non-linear} multigrid method (see Hackbush, 1985) to multilevel RT. 

The iterative scheme of the non-linear multigrid method
is composed of two parts:
a {\it smoothing} one where a small number of 
GS-based iterations on the desired
{\it finest} grid are used to get rid of the high-frequency
spatial components of the error in the current estimate,
and a correction obtained from the solution of an 
error equation in a {\it coarser} grid. 
With our non-linear multigrid code the total computational work scales simply as NP,
although it must be said that
the computing time per iteration is about 4 times larger than that
required by the $\Lambda-$iteration method.

\begin{figure}[ht]
  \begin{center}
    \epsfig{file=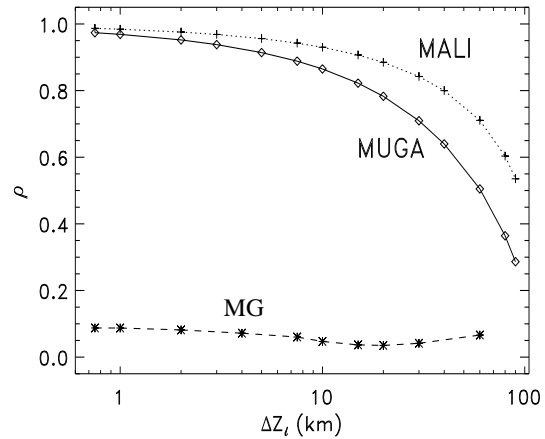, width=8cm}
  \end{center}
\caption{Variation with the grid-spacing $\Delta{z}$
of the maximum eigenvalue of the iteration
operator corresponding to several multilevel iterative schemes.
The MG symbol refers to our {\it non-linear} multigrid code,
while MUGA to our multilevel GS-based code. MALI
refers to the Jacobi-based multilevel ALI method of Rybicki \& Hummer, (1991).}  
\label{fig5}
\end{figure}

In order to compare the convergence rate of all these 
iterative methods we
present in Fig. 2 an estimate of the maximum eigenvalues ($\rho$) of the
corresponding iteration operator,
which controls the convergence properties
of such iterative schemes.
The knowledge of this maximum eigenvalue ($\rho$) is useful because
errors decrease as ${\rho}^{itr}$, 
where ``$itr$'' is the iterative step. As it can be noted in Fig. 2 the
convergence rate of both, the MALI and MUGA schemes decreases when the spatial
resolution of the grid is improved, while the maximum eigenvalue
of our {\it non-linear} multigrid method is always 
very small ($\rho\sim0.1$) and
{\it insensitive} to the grid-size. A maximum eigenvalue $\rho=0.1$ means that
the error decreases by one order of magnitude each time we perform an iteration!
This explains that, typically, two multigrid iterations
are sufficient to reach the self-consistent solution for the atomic level populations.

\section{Formal solvers for RT applications}

The formal solution routines of our NLTE codes
(for unpolarized or polarized radiation and
for atomic or molecular species) are based on improvements and
generalizations of the {\it short-characteristics} (SC)
technique introduced by Kunasz \& Auer (1988). 
Let us recall it briefly
indicating also our generalization to 3D radiative transfer
and to the case of polarized radiation.

The scalar RT equation for the specific intensity is

\begin{equation}
{{{\rm d}{\rm I}_{\nu}}\over{{\rm d}{s}}}\,=\,
{\chi_\nu}\,(\,{\rm S}_{\nu}\,-\,{\rm I}_{\nu}\,)\,, 
\end{equation}
where $s$ is the geometric distance along the ray propagating
in a certain direction
in a 3D medium, $\chi_\nu$ is the total opacity
and ${\rm S}_{\nu}$ the source function. 

Point O is the grid-point 
of interest at which one wishes to calculate
the specific intensity ${\rm I}_{\rm O}$, for a given frequency ($\nu$) 
and a direction (${\vec{\Omega}}$).
Point M is the the intersection point
with the grid-plane that one finds when moving along $-\vec{\Omega}$.
At this {\it upwind} point the specific intensity
${\rm I}_{\rm M}$ (for the same frequency and angle) is known
from previous steps. In a similar way, point
P is the intersection point with the grid-plane
that one encounters when moving along 
$\vec{\Omega}$. We also introduce the optical depths
along the ray between points M and O ($\Delta{\tau}_{\rm M}$)
and between points O and P ($\Delta{\tau}_{\rm P}$). 
From the formal solution of the previous
transfer equation one finds that

\begin{equation}
{\rm I}_{\rm O}={\rm I}_{\rm M}\,{\rm e}^{-\,\Delta{\tau}_{\rm M}}\,+\,
\int_{0}^{\Delta{\tau}_{\rm M}}\,{\rm S}(t)
{\rm e}^{-({\Delta{\tau}_{\rm M}}\,-\,t)}\,dt,
\end{equation}
with the optical depth variable measured
from M to O.

The integral of this equation can be solved {\it analytically} by
integrating along the {\it short-characteristics} MO assuming
that the source function ${\rm S}(t)$ varies {\it parabolically}
along M,O and P. The result reads:

\begin{equation}
{\rm I}_{\rm O}={\rm I}_{\rm M}\,{\rm e}^{-\Delta{\tau}_{\rm M}}\,+\,
{\Psi}_{\rm M}{\rm S}_{\rm M}\,+\,{\Psi}_{\rm O}{\rm S}_{\rm O}\,+\,
{\Psi}_{\rm P}{\rm S}_{\rm P},
\end{equation}
where $\Psi_{\rm X}$ (with X either M, O or P)
are given in terms of the quantities $\Delta{\tau}_{\rm M}$ and
$\Delta{\tau}_{\rm P}$ that we evaluate numerically by assuming
that ${\rm ln}(\chi)$ varies linearly with the geometrical depth, $\chi$ 
being the opacity.


If the interest lies in the generation and transfer of {\it polarized}
radiation in magnetized astrophysical plasmas
(cf. Trujillo Bueno and Landi Degl'Innocenti, 1996; Trujillo Bueno, 1999; 2001)
the situation is a bit more complicated because, instead of having to solve the previous
RT equation for the specific intensity, one has to solve, in general, a vectorial
transfer equation for the {\it four} Stokes parameters. For example, for the
standard case of polarization induced by the Zeeman effect, 
the Stokes-vector at the grid-point O is

\begin{equation}
{\bf I}_{\rm O}={\bf O}(0,\Delta{\tau}_{\rm M}){\bf I}_{\rm M}\,+\,
\int_{0}^{\Delta{\tau}_{\rm M}}\,{\bf O}(t,\Delta{\tau}_{\rm M})\,{\bf S}(t)\,dt,
\end{equation}
where ${\bf O}(t,\Delta{\tau}_{\rm M})$ is the evolution operator
(i.e. the 4$\times$4 Mueller matrix of the atmospheric slab between $t$
and $\Delta{\tau}_{\rm M}$). In general, this evolution operator
does not have an easy analytical expression and the integral of the previous
equation cannot be solved analytically. However, if the 4$\times$4
absorption matrix conmutes between depth points M and O
(e.g. because one assumes the absorption matrix to be constant between
M and O and equal to its true value at the middle point)
the evolution operator reduces then to an expression given by the exponential 
of the absorption matrix. The integral of Eq. (15)
can then be solved analytically provided that the source function vector
$\bf S$ is assumed to vary parabolically along M, O and P. Our formal
solution method for Zeeman line transfer
is precisely based on this idea and the results reads:

\begin{equation}
{\bf I}_{\rm O}={\bf O}(0,\Delta{\tau}_{\rm M}){\bf I}_{\rm M}\,+\,
{\bf \Psi}_{\rm M}{\bf S}_{\rm M}\,+\,{\bf \Psi}_{\rm O}{\bf S}_{\rm O}\,+\,
{\bf \Psi}_{\rm P}{\bf S}_{\rm P},
\end{equation}
where $\bf \Psi_{\rm X}$ (with X either M, O or P) are 4$\times$4
matrices which are given in terms of $\Delta{\tau}_{\rm M}$ and
$\Delta{\tau}_{\rm P}$, in terms of the inverse of the absorption matrix
and in terms of the analytical expression given by
Landi Degl'Innocenti and Landi Degl'Innocenti (1985)
for the evolution operator for the case
in which the absorption matrix is assumed to be constant
between the spatial grid points. An alternative formal solver
of the Stokes-vector transfer equation suitable for NLTE applications 
is the one given by
Eqs. (1-4) of Socas-Navarro, Trujillo Bueno \& Ruiz Cobo (2000).

The application of these formal solution methods in 1D is straightforward.
The generalization to
2D geometries with horizontal periodic boundary conditions is 
substantially more complicated. A suitable strategy has been described by 
Auer, Fabiani Bendicho and Trujillo Bueno (1994).

The main changes when going to 3D 
imposing horizontal periodic boundary conditions
lie in the interpolation. We have
assumed that $\rm I_M$ is known but, in most cases,
the M-point (like the point P)
will not be a grid-point of the chosen 3D spatial grid. The
intensity at this M-point has to be calculated
by interpolating from the available information at the nine surrounding
grid-points, as we must also do for obtaining the opacities
and source functions at M and P.
Parabolic interpolation can however generate spurious
negative intensities if the spatial variation of the physical quantities
is not well resolved by the spatial grid. This happens,
for instance, if one tries to 
simulate the propagation of a beam in
vacuum using a three dimensional grid.  
To avoid these problems we have improved the 1D monotonic interpolation
strategy of Auer and Paletou (1994), and generalized it 
to the two-dimensional parabolic interpolation that is required for
3D RT calculations with multilevel atomic models 
(see Fabiani Bendicho \& Trujillo Bueno, 1999).


\section{Conclusions}
\label{conclus}

The RT methods presented here are especially attractive
because of their direct applicability to a variety of
complicated RT problems of astrophysical interest.
We emphasize that their convergence rate are extremely high, that
they do not require the construction and the inversion of any large matrix
and that the computing time per iteration is very small.

\end{document}